\documentclass[aps,prb,twocolumn,showpacs]{revtex4}
\usepackage{amsmath}
\usepackage{dcolumn}
\usepackage{graphicx}
\usepackage{float}

\begin{document}

\title{Collective vortex pinning and crossover between second order to first order transition in optimally doped Ba$_{1-x}$K$_x$BiO$_3$ single crystals}
\author{Yanjing Jiao, Wang Cheng, Qiang Deng, Huan Yang and Hai-Hu Wen*\email}
\affiliation{National Laboratory of Solid State Microstructures and Department of Physics,
Collaborative Innovation Center of Advanced Microstructures, Nanjing University, Nanjing 210093, China}

\begin{abstract}
Measurements on magnetization and relaxation have been carried out on an optimally doped Ba$_{0.59}$K$_{0.47}$BiO$_{3+\delta}$ single crystal
with $T_c$ = 31.3 K. Detailed analysis is undertaken on the data. Both the dynamical relaxation and conventional relaxation
have been measured leading to the self-consistent determination of the magnetization relaxation rate. It is found that the data are
well described by the collective pinning model leading to the glassy exponent of about $\mu\approx$ 1.64 to 1.68 with the magnetic fields of 1 and 3 T.
The analysis based on Maley's method combining with the conventional relaxation data allows us to determine the current dependent activation
energy $U$ which yields a $\mu$ value of about 1.23 to 1.29 for the magnetic fields of 1 and 3 T. The second magnetization peaks appear in wide temperature region from 2 K to 24 K.
The separation between the second peak field and the irreversibility field becomes narrow when temperature is increased.
When the two fields are close to each other, we find that the second peak evolves into a step like transition of magnetization, suggesting a crossover from
the second order to first order transition. Finally, we present a vortex phase diagram and demonstrate that the vortex dynamics
in Ba$_{1-x}$K$_x$BiO$_3$ can be used as a model system for studying the collective vortex pining.
\end{abstract}


\maketitle

\section{Introduction}

Vortex related properties are very interesting and important for type-II superconductors, especially for the high power applications.
Numerous detailed studies on the vortex dynamics of cuprate and iron based superconductors have been performed\cite{BlatterRMP,WenPhysicaC,Krusin,Klein}.
Some interesting phenomena like magnetic flux jump\cite{PRB72,PRB73}, collective vortex creep\cite{PRB70},
and second peak effect have been discovered through the experiments.
Although the collective vortex creep theory has successfully interpreted some features or vortex dynamics\cite{PRBThompson,PRB62,YangHPRB,NPRB84,TPRB86} in cuprate and iron based superconducting materials, the crystal structures in those systems are generally complicated, which will bring in the complex in the behavior of vortex physics. For example, a quite sophisticated  vortex phase diagram has been reported for some cuprate superconductors, such as the Bi$_2$Sr$_2$CaCu$_2$O$_8$ (Bi2212)\cite{Bi2212}. So the research of vortex physics in crystals with simpler structure may give us a better way to understand
vortex motion and pinning mechanism, especially for checking the collective vortex pinning. For this reason, the Ba$_{1-x}$K$_x$BiO$_3$ (BKBO) superconductor has attracted much attention after the discovery\cite{Mattheiss,Cavanature}. It has relatively simple basic cubic perovskite crystal structure
for $0.375 < x < 0.5$ and it is close to a three dimensional system\cite{PRB41}. Moreover, the coherence length in BKBO is about 3-6 nm \cite{PRB49,PRB40}.
This value is higher than that of cuprate superconductors, which may suggest that the weak link effect between grains is not very crucial and the simple powder-in-tube technique for making the superconducting wires can be applied. This long coherence length is also helpful for making the superconducting quantum coherence device. In addition, in the BKBO superconductor, the conducting layer is BiO layer. Compared with the cuprate superconductors,
the most conspicuous feature of BKBO is the lack of two-dimensional CuO$_2$ planes. So the unconventional, most likely the magnetic origin driven mechanism proposed for cuprate superconductors may not be suitable for BKBO\cite{PRB58}. Some other anomalous features also have been discovered from the experiments, including a large oxygen isotope effect\cite{nature335,prl66} and the softening of phonons in the BKBO system \cite{Js8,phyC471}.
The second magnetization peak effect has been observed in the underdoped Ba$_{0.66}$K$_{0.32}$BiO$_{3+\delta}$ with $T_c$ = 27 K\cite{TaojianPRB}.
In that paper they observed second magnetization peak in the intermediate temperature region and its general behavior looks
like that in YBa$_2$Cu$_3$O$_7$ (YBCO). But vortex dynamics on optimally doped BKBO single crystal with a higher $T_c$ K has
not been studied before. In present work, we make a detailed study on the vortex dynamics of optimally doped BKBO single crystal with measurements
of magnetization and relaxation, including the conventional and dynamical relaxations.

\section{Experiment}

Near optimally doped single crystals of Ba$_{1-x}$K$_x$BiO$_3$ are grown by the molten-salt-electrochemical
method which is similar to that reported previously \cite{Norton,Nishio357}. Molten KOH with slight deionized water were used as the growth environment. We adjust the mass ratio between Ba(OH)${_2}\cdot8$H${_2}$O
and Bi${_2}$O${_3}$ to about 2:5 to grow the optimally doped sample, and the growth time is about 3 days. With this method,
we successfully get single crystals with volume in the range of 2-5 mm$^{3}$. To check the quality of these single crystals,
X-ray diffraction (XRD) measurements are performed on the samples by a Bruker D8 Advanced diffractometer with the Cu-K radiation.
The XRD pattern of the sample matches very well with the previous reported results in the slightly underdoped samples\cite{TaojianPRB}.
The chemical composition of the sample was characterized by using the Energy-dispersive X-ray spectroscopy (EDX/EDS)
attached to a scanning electron
microscope (Hitachi Co., Ltd.), and the resultant chemical composition is close to Ba$_{0.59}$K$_{0.47}$BiO$_{3+\delta}$, which is close to the reported optimal doping value x $\approx$ 0.4 to 0.5.
All the magnetization measurements were carried out with a SQUID magnetometer ( SQUID-VSM-7T, Quantum Design). During the measurements,
the magnetic field $H$ was always parallel to $c$ axis of the sample. Resistivity was measured by the standard four-lead method
in a Quantum Design instrument, the physical property measurement system (PPMS).

\section{Results and Discussions}
\subsection{Superconducting transition and second peak effect}

\begin{figure}
\includegraphics[width=8.5cm]{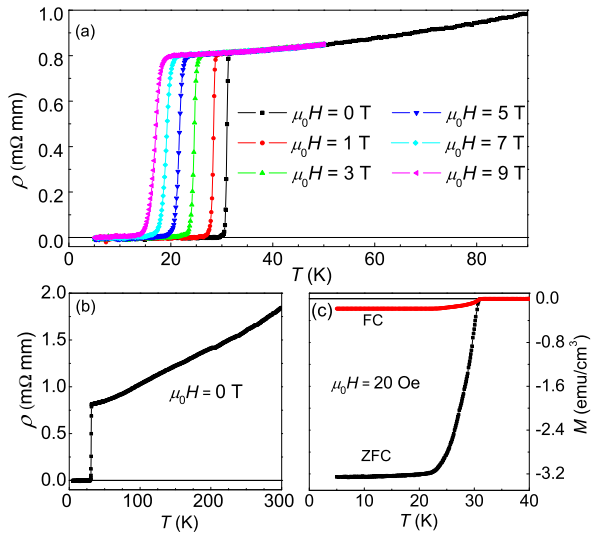}
\caption{(Color online) (a) Temperature dependence of resistivity of Ba$_{0.59}$K$_{0.47}$BiO$_{3+\delta}$ measured
at various magnetic fields. (b) Temperature dependent resistivity at zero field with the temperature range
from 2 K to room temperature. (c) Magnetization measured in the ZFC and FC processes at a magnetic field of 20 Oe.}\label{fig1}
\end{figure}

Figure~\ref{fig1}(a) shows the temperature dependent resistivity $\rho$ at different magnetic fields ranging
from 0 to 9 T. The critical temperature $T_{c0}$ taken from zero resistance at zero field is about 30 K,
and the onset critical temperature $T_c^{onset}$ with a criterion of $90\%\rho_n$ is 31.3 K, here $\rho_n$ is the normal state resistivity. The narrow transition width shows the high quality of the sample. Figure~\ref{fig1}(b) is the $\rho$-$T$ curve measured at zero magnetic field
with a wide temperature range from 2 K to room temperature. The residual resistance ratio $RRR=R(T=300\ \mathrm{K})/R(T=32\ K)=2.28$,
which is rather small and indicates the strong scattering of the charge carriers in the sample. Temperature dependence of magnetization measured
with zero-field-cooling (ZFC) and field-cooling (FC) process at 20 Oe is shown in Fig.~\ref{fig1}(c). The irreversibility
temperature of the ZFC-FC magnetization curves at 20 Oe is about 31.1 K which is comparable to $T_{c0}$ from the resistance measurements.
The magnetization measured in the ZFC process in the temperature region from 2 to 20 K shows a flat temperature dependence, which indicates again a full magnetic shielding effect. The large difference between the ZFC and FC magnetization curves indicates a strong hysteresis of magnetization and thus strong vortex pinning.

\begin{figure}
\includegraphics[width=8.5cm]{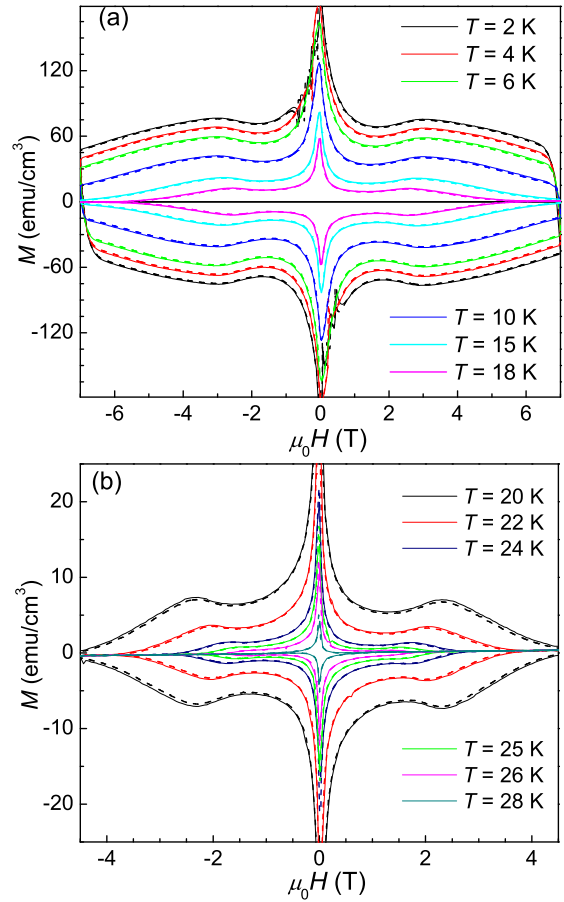}
\caption{(Color online) Magnetization hysteresis loops of the BKBO single crystal at various temperatures
ranging from (a) 2 K to 18 K, (b) 20 K to 28 K under different field sweeping rates of 200 Oe/s (solid lines) and 50 Oe/s
 (dashed lines), respectively.}\label{fig2}
\end{figure}

Figure~\ref{fig2} shows the magnetization hysteresis loops (MHLs) at different temperatures
from 2 K to 28 K with different magnetic field sweeping rates. The symmetric MHLs indicate that the bulk vortex pinning rather than the surface shielding current dominates the magnetization in the sample. We have measured the MHLs with two different field sweeping rates: 50 Oe/s and 200 Oe/s. One can see clear difference between the MHLs shown by solid ($dH/dt$ = 200 Oe/s) and dashed ($dH/dt$ = 50 Oe/s) line. This difference will allow us to extract the dynamical relaxation rate. From Fig.~\ref{fig2},
 one can find obvious second peak effect over a wide temperature range beginning from the lowest temperature 2 K.
 The so-called second peak of magnetization is relative to the first one near the zero magnetic field. It should be noted that flux jump effect shows up as the large scale anomalous jumps of the magnetization below 4 K in the low field range, which is similar to the result in the underdoped sample\cite{TaojianPRB}.

\begin{figure}
\includegraphics[width=8.5cm]{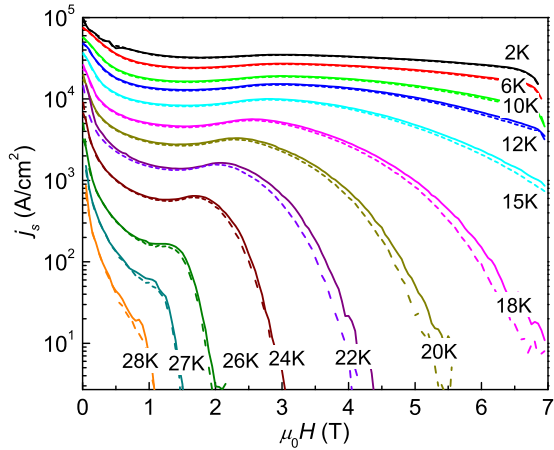}
\caption{(Color online) Magnetic field dependence of $j_s$ calculated from the Bean critical state model at different
temperatures in semi-log plot. The solid lines stand for the $j_s$ with field sweeping rate of 200 Oe/s,
while the dashed lines represent those with 50 Oe/s.}\label{fig3}
\end{figure}

\subsection{Theoretical models for treating the data}
The vortex dynamics is mainly described with the model of thermally activated flux motion (TAFM)\cite{Anderson}\label{eq1}. Based on this model, the electric field $E$ induced by the vortex motion is written as

\begin{equation}
 E=v_0B \times \exp [\frac{-U(j_s)}{k_BT}].\label{eqTAFM}
\end{equation}

In above equation, $v_0$ is the largest vortex moving speed, $B$ is the averaged magnetic induction, $U(j_s)$ is the flux activation energy which is defined as a function of current density $j_s$. Generally, it is also a function of temperature and magnetic field. Thus a general description of the activation energy was suggested as\cite{PRBFeigelman,Malozemoff}

\begin{equation}
U(j_s)=\frac{U_c}{\mu}[(\frac{j_c}{j_s})^{\mu}-1],\label{eq1}
\end{equation}

where $U_c$ is the intrinsic pinning energy, $j_c$ is the unrelaxed critical
current density which is always larger than transient current density $j_s$, $\mu$ is the glassy exponent that involves the most attempting size
of the vortex motion and the wandering factor of an elastic object in a three dimensional manifold\cite{VinokurPRL,BlatterRMP}. In a three-dimensional system $\mu \textgreater$ $0$ can be
used to describe a elastic flux creep while $\mu \textless$ $0$ for plastic motion\cite{WenPRL1997,WenPhysicaC1998}. In particular, $\mu = -1$ corresponds to the
classical Kim-Anderson model\cite{KimAnderson}.

Combining above two equations, we can derive

\begin{equation}
j_s=j_c[1+\frac{\mu k_BT}{U_c}\ln(\frac{v_0B}{E})]^{-1/\mu}.\label{eqjs}
\end{equation}

We can also further understand the second peak effect from field dependent transient superconducting current
density $j_s$-$\mu_cH$ curve. The $j_s$ can be obtained based on the extended Bean critical state model\cite{Bean},
\begin{equation}
j_s=20\frac{\Delta M}{a(1-a/3b)}.
\end{equation}
Here the field dependent irreversible magnetization width $\Delta M$, which is corresponding to the transient critical current density in
a superconductor, and can be calculated by $\Delta M =M_{in}-M_{de}$ . Here $M_{in}$ ($M_{de}$) is the magnetization associated with increasing
(decreasing) magnetic fields with the unit of emu/cm$^3$ . The length scales $a$ and $b$ in unit of cm are the width and length of the sample ($a<b$).
In Fig.~\ref{fig3}, we show field dependent $j_s$ with two different field sweeping rates of 200 Oe/s and 50 Oe/s. It is obvious
that $j_s$ are different from each other with different field sweeping rates, a faster sweeping rate corresponds to a higher current density.
The calculated $j_s$ at 2 K and 0 T can reach a value of $10^5$ A/cm$^2$. Moreover, the second peak effect is very clear in Fig.~\ref{fig3}, i.e., at certain temperature, curve $j_s$ decreases first, then increases, and finally keeps decreasing with the increase of magnetic field.  The general shape of magnetization hysteresis
loops and related features are similar to that in cuprate superconductor YBCO\cite{Kokkaliaris}.
In addition, the peak position $H_{sp}$ moves closer to the irreversibility field $H_{irr}$ with increasing of temperature resulting in
a continuous increasing ratio of $H_{sp}/H_{irr}$. The origin of this second peak effect has several possible scenarios. On picture explains this effect as the crossover between different vortex pinning regimes. The dominated single-vortex pinning at low magnetic field may be replaced by pinning of vortex bundles or entangled vortices in high
field region, and the transient superconducting current keeps increasing with the magnetic field as the flux creep is lowered. However, in high magnetic region, some dislocations will be formed in the vortex system leading to the plastic vortex motion, the transient critical current density $j_s$ will decrease. At higher
temperatures, the second peak effect becomes weaker and $H_{sp}$ becomes more closer to the $H_{irr}$. From our data, we can see that the second peak effect becomes a step like transition when temperature is 28 K.

\subsection{Magnetization relaxation rate and glassy exponent}

\begin{figure}
\includegraphics[width=8.5cm]{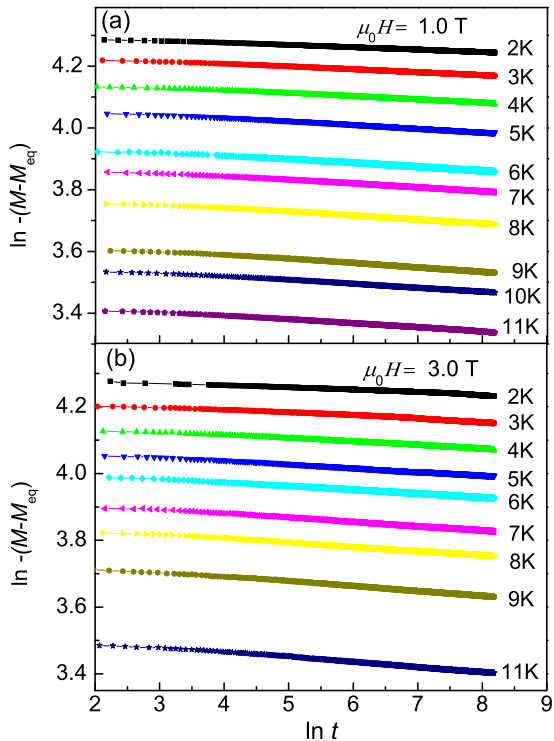}
\caption{(Color online) Time dependence of the nonequilibrium magnetization on a log-log plot at various temperatures
at fields of (a) 1 T (b) 3 T. Here the magnetic field is applied parallel to $c$-axis of the sample.}\label{fig4}
\end{figure}

It has been proven that magnetization relaxation is a very effective way to investigate the vortex dynamics\cite{Brandt}.
The magnetization relaxation can be categorized into the conventional relaxation and dynamical relaxation. In a typical conventional magnetic relaxation measurement, time dependent decay of magnetization is measured after applying a field following the ZFC process. The corresponding magnetization-relaxation rate $S$ is defined as

\begin{equation}
S=-\frac {d\ln[-(M-M_{eq})]}{d\ln t}.\label{eqS}
\end{equation}

Here $M$ is the instantaneously measured magnetization, and $M_{eq}$ is the equilibrium magnetization which is taken from the average of magnetization
of increasing and decreasing field of the full MHL taken at a certain magnetic field and corresponding temperature. For data measured at high temperatures, it is very important to remove this equilibrium magnetization away since the MHL becomes asymmetric. The time $t$ is taken from
the moment when the critical state is prepared. A long waiting time is necessary for determining the meaningful relaxation rate.

In contrast, the so called dynamic magnetization-relaxation measurements are carried out by measuring the MHLs with different field sweeping rates $dH/dt$
as described above. The corresponding magnetization-relaxation rate $Q$ can be obtained from

\begin{equation}
Q=\frac{d\ln j_s}{d\ln(dB/dt)}.\label{eqQ}
\end{equation}

The very small difference between the two $j_s$-$\mu_cH$ curves measured with four times difference of sweeping rate
corresponds to a very small magnitude of $Q$. From Eq.3, one can easily have the message that $S \approx Q$. Here, in deriving the conventional magnetization relaxation rate, we assume that $E \propto dj_s/dt$; in deriving the dynamical relaxation rate, we assume $E\propto dB/dt$.

\begin{figure}
\includegraphics[width=8.5cm]{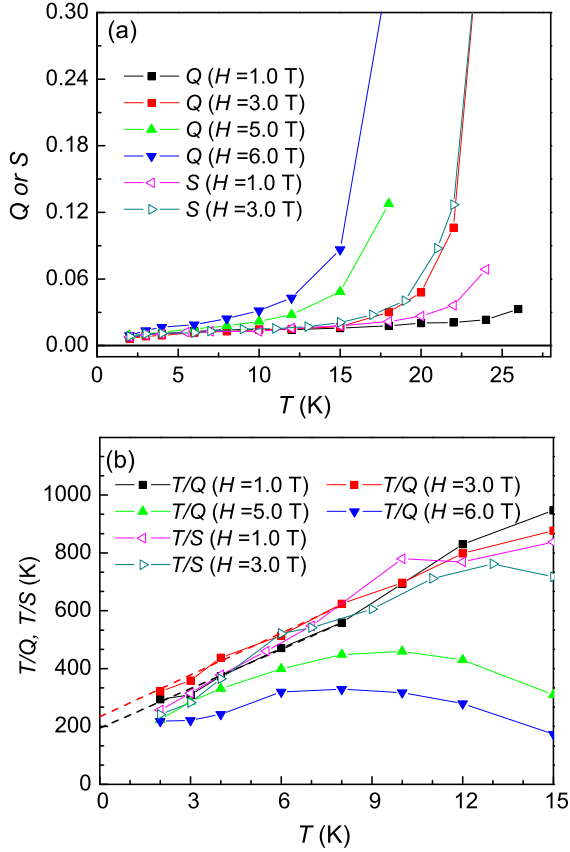}
\caption{(Color online) (a) Temperature dependence of the dynamic relaxation rate $Q$ ranging from 1 T to 6 T,
and the conventional relaxation rate $S$ at 1 T and 3 T. (b) Temperature dependence of the ratio $T/Q$ and $T/S$
at various magnetic fields.}\label{fig5}
\end{figure}

The time dependence of magnetization at various temperatures are shown in Fig.~\ref{fig4} at magnetic fields of 1 T and 3 T.
Here almost linear relationship between $-(M-M_{eq})$ and time in log-log plot is observed in the long time limit, being consistent with the  model of thermally
activated flux motion. According to the definition of the conventional relaxation rate as Eq.~\ref{eqS}, the absolute value of the slope of the log-log plot in Fig.~\ref{fig4} gives the conventional relaxation rate, we thus determine $S$ through a linear fit to the double logarithmic plot from the data in the long time limit with the time window of $lnt$ = 4 to 8.18. The dynamic relaxation rates $Q$ calculated from Eq.~\ref{eqQ} and Fig.~\ref{fig3}
as well as the conventional relaxation rates $S$ calculated from Eq.~\ref{eqS} and Fig.~\ref{fig4} versus temperature are shown in Fig.~\ref{fig5}(a).
One can find that relaxation rates calculated from the two methods at the same field are consistent with each other.
To get more information about the vortex dynamics, we can calculate the activation energy and the glassy exponent which is strongly related to the relaxation
rate. If combining Eq.~\ref{eqTAFM} and Eq.~\ref{eqjs} and the definition of the dynamic relaxation mentioned above, the following equation
can be obtained\cite{WenAlloys},

\begin{equation}
\frac{T}{Q(T)}=\frac{U_c(T)}{k_B}+\mu CT,\label{eqTQ}
\end{equation}

where $C=\ln(v_0B/E)$ is a parameter that almost independent of temperature. According to this equation, the
relaxation rate $Q$ is determined by the balance between the two temperature dependent parts $U_c(T)$ and $\mu CT$, and shows a complex temperature-dependent behavior.
However if the second part is too small and can be ignored at very low temperature, then $Q$ will show a linear dependence on $T$, the ratio $T/Q$ should give a constant which is just the intrinsic pinning energy.
By extrapolating the curve $T/Q$ down to zero temperature, one can obtain the value of $U_c(0)$. The value of $\mu C$ can be determined
from the slope of $T/Q$ vs. $T$. Meanwhile, from Eq.\ref{eqjs} one can derive that

\begin{equation}
-d\ln j_s/dT=C\times\frac{Q(T)}{T}.\label{eqC}
\end{equation}

\begin{figure}
\includegraphics[width=8.5cm]{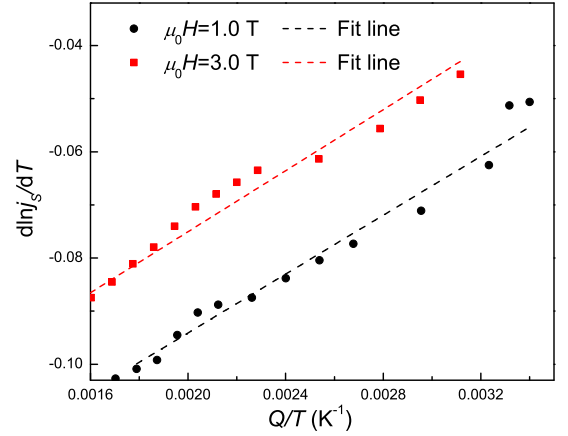}
\caption{(Color online) Correlation between $d\ln j_s(T)/dT$ vs. $Q(T)/T$ at low temperatures for the magnetic fields of 1 and 3 T. In order to calculate the value, we interpolate the data of $\ln j_s$ vs. $T$, and $Q(T)/T$ vs. $T$ between 2 K and about 8 K, then get the corresponding data at the same temperature from both correlations and plot them here. The slope gives the value of C. }\label{fig6}
\end{figure}

Therefore the parameter $C$ can be calculated\cite{WenPhysicaC} by the slop of $-d\ln j_s/dT$ vs. $Q/T$. In Fig.\ref{fig6}, we show the data of -$\ln j_s(T)$ vs. $Q(T)/T$ at low temperatures for the magnetic field of 1 and 3 T. One can see that the curves are roughly straight, so we can get the $C$ values of 27.6 and 28.7 for 1 and 3 T, respectively. With the values of $\mu C$ determined above, we can determine the glassy exponent $\mu$. We plot temperature dependence of $T/Q$ and $T/S$ at different magnetic fields in Fig.~\ref{fig5}(b). Here we do a linear fit of the data $T/Q$ vs. $T$ between 2 and 8 K to get the intercept at $T$ = 0 K and the slope.
From the intercept value at $T$ = 0 K, we get the the values of $U_c(0)/k_B$ = 194 K and 233 K at 1 T and 3 T, respectively. These values are larger than that of $Ba$(Fe$_{1-x}$Co$_x$)$_2$As$_2$ single crystals\cite{ShenPRB}. Based on the method that we discussed above, the values of $\mu C\approx45.38$ and $48.18$ at 1 T and 3T can
be determined. Meanwhile, taking the values of $C$ that determined above, we can get the glassy exponent
$\mu\approx$ 1.64 at 1 T and 1.68 at 3 T.

The relaxation rates show a monotonic temperature dependence for different fields, as shown in Fig.~\ref{fig5}(a). While it exhibits clearly a plateau
in the intermediate temperature region, which was also observed in YBa$_2$Cu$_3$O$_{7-\delta}$\cite{RMP68}. The temperature window of
the plateau in $Q$-$T$ curve shrinks when magnetic field is increased and almost disappears at $\mu_{0}H$ = 6.0 T. This plateau can
be explained by the vortex collective pinning model as Eq.~\ref{eqTQ}. When the second term $\mu CT$ is much larger than $U_c$ at high
temperatures, the first term can be ignored and almost temperature independent value of $Q$ can be obtained. This plateau of $S$ or $Q$ in the intermediate temperature region strongly suggests the applicability of the collective vortex pinning model.

\begin{figure}
\includegraphics[width=8.5cm]{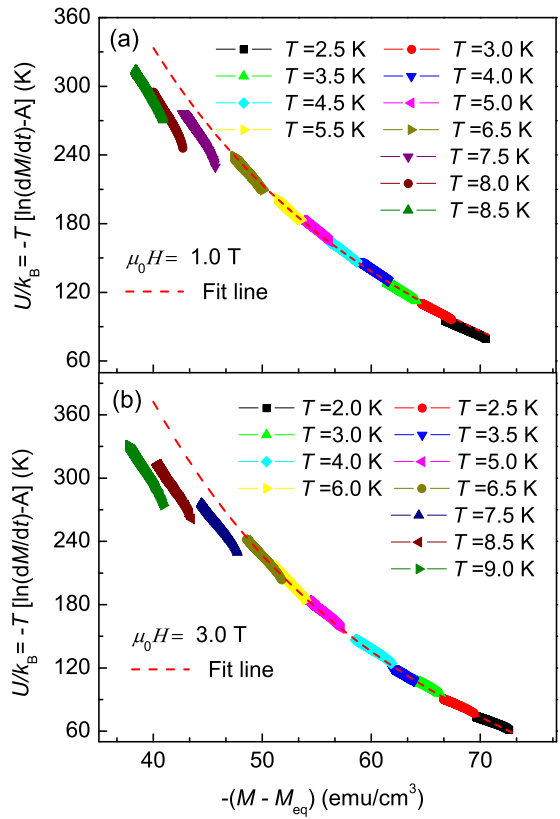}
\caption{(Color online) Nonequilibrium magnetization dependence of the scaled activation energy $U$ obtained by the
Maley¡¯s method with parameter $A$ = 28. A fitting result with Eq.~\ref{eq1} is presented as the red dot line with the parameter $\mu$ = 1.23 with a magnetic field of 1 T. (b) The same as (a)
while with a different magnetic field of 3 T, the parameters $A$ = 28, the glassy exponent $\mu$ = 1.29.}\label{fig7}
\end{figure}

\subsection{To derive the $U(j_s)$ dependence and $\mu$ parameter from Maley's method}

In order to verify the calculated activation energy and glassy exponent, we also use the Maley's method\cite{Maley}
to analyze the data of time dependent nonequilibrium magnetization. From Eq.~\ref{eqTAFM}. and the basic assumption of $E\propto |dM/dt|$, we have
\begin{equation}
\frac {U(j_s)}{\textrm{$k_B$}}=-T[{\ln(\textrm{d}M/\textrm{d}t)}-A],
\end{equation}
where $A$ is a time-independent constant related to the average hoping velocity. By adjusting $A$ we can scale the curves
measured at low but different temperatures onto one curve. This is shown in Fig.~\ref{fig7}. From the figure, we can see that all the curves
can be well scaled together at temperatures below 6.5 K, while the scaling fails at higher temperatures.
The temperature region for the good scaling may be in the range where the temperature dependence of $U_c(T)$ is weak.
The glassy exponent can be obtained by fitting the scaled curves with Eq.~\ref{eq1} by TAFM model. The fitting gives
an independent evaluation of  $U_c/k_B$ = 199 K, $\mu$ = 1.23 at 1 T, and $U_c/k_B$ = 269 K, $\mu$ = 1.29 at 3 T, which are close
to the results from analysis based on Eq.~\ref{eqTQ}. The difference between the $\mu$ values derived in the two methods is induced by the distinct and crude assumptions adopted in analyzing the data. Therefore we can conclude that the $\mu$ value may locate in the region of 1.2 to 1.7. As we discussed above, in the model of collective creep, the glassy exponent $\mu$ influences
on the vortex dynamics. Actually, theoretical prediction\cite{VinokurPRL} for the $\mu$ values is that: $\mu$ = 1/7 for single vortex; $\mu$ = 3/2 for small bundles of vortices; $\mu$ = 7/9 for large bundles of vortices. The value $\mu$ = 3/2 locates just
between the two calculated values that obtained by different methods. So our results here may suggest that the collective vortex motion is in the small bundles vortex regime. Nevertheless, the $\mu$ values derived in different methods suggest that the collective pinning model is applicable in this kind of superconductors.

\subsection{Crossover from the second to first order transition at high temperatures}
\begin{figure}
\includegraphics[width=8.5cm]{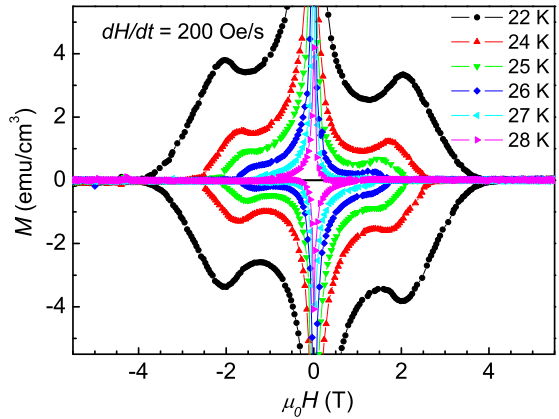}
\caption{(Color online) Enlarged view of the MHLs measured at temperatures 22K 24K 25K 26K 27K and 28K from outside to inside. With increasing temperature, the MHL near the irreversibility point $H_{irr}$ changes from a smooth long tail like transition to a step like transition. }\label{fig8}
\end{figure}

Recalling the MHLs with the second peak effect, we see that there is a large gap between the second peak field $H_{sp}$ and the irreversible field $H_{irr}$ when the temperature is not high. In this case, the merging of the magnetization curves in the field ascending and descending processes is gradual, showing a long tail of the critical current density $j_s$ versus $H$, as shown in Fig. 3 for $T$ = 18 K to 24 K. However, when the temperature is increased further, this gap becomes narrow and narrow, showing a step like behavior in both the MHL near the $H_{irr}$ or the curve of $j_s$ versus $H$. We present the MHLs at high temperatures (22 to 28 K) in Fig.~\ref{fig8}. It is clear that the MHLs near the irreversibility line is indeed changing from a long tail behavior to a step like. We argue this change of the behavior from a smooth to a step like transition as the crossover from the second order transition of vortices to the first order transition. The second order transition near $H_{irr}$ at a low temperature and high magnetic field is understood as the transition from a plastic motion dominated vortex dynamics to the vortex liquid. While the step like transition corresponds to a more ordered vortex system (now the vortex density is quite low) to the vortex liquid transition. Our observation here is quite close to the theoretically predicted phase diagram\cite{BlatterRMP} in the high temperature and low field region, where the first order transition occurs and the phase line connects to the second order transition at high magnetic fields. This first order transition may be similar to that observed in YBCO single crystals\cite{Kwok2,Kwok}, but the magnetic field is much lower in present BKBO sample. Since our sample is a bulk, the global measurement for the step like transition gives still a certain width of the transition. It is highly desired to use the small Hall probe or local measurements to detect this first order transition as that done in cuprate superconductor Bi2212\cite{Bi2212,Zeldov,Yeshurun}.

\subsection{Vortex phase diagram and general discussion}

\begin{figure}
\includegraphics[width=8.5cm]{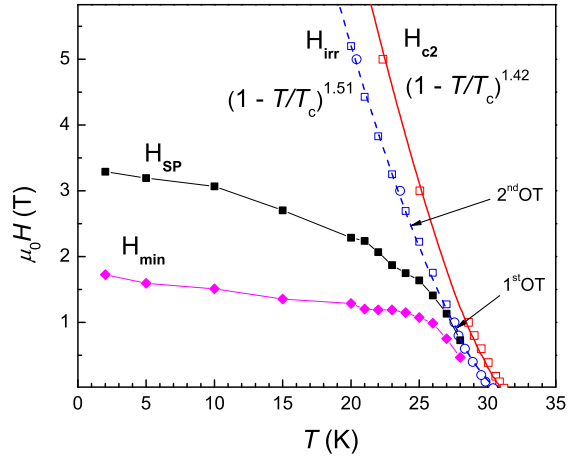}
\caption{(Color online) The vortex phase diagram of the optimally doped sample BKBO.
The filled symbols are taken from the magnetization measurements while the open ones are taken from the resistive measurements. The upper critical field $H_{c2}$ is determined with the criterion of 90\%$\rho_n$, while the irreversibility field $H_{c2}$ is determined with 10\%$\rho_n$.
The red solid ($H_{c2}$) and blue dashed ($H_{irr}$) lines are the fitting curves with the formula $H(T)=H(0)(1-T/T_c)^n$.
}\label{fig9}
\end{figure}

With above results, finally, we depict a vortex phase diagram for the optimally doped
Ba$_{0.59}$K$_{0.47}$BiO$_{3+\delta}$ single crystal in Fig.~\ref{fig9}. Here $H_{irr}$ is the irreversibility field that is
determined from $j_s$-$\mu_0H$ curves with the criterion of $j_s$ = 10 A/cm$^2$ (blue open circles). The field position $H_{min}$ is the point with minimum $j_s$ between the first and the second magnetization peak position $H_{SP}$. The upper critical field $H_{c2}$ and the irreversible
field $H_{irr}$ shown as open squares are obtained from the resistive measurement with criterions of $90\%\rho_n$ and of $10\%\rho_n$, respectively.
Below the second peak effect field, the vortex motion can be well interpreted with the collective vortex motion with the glassy exponent $\mu\approx$ 1.2 to 1.7, this suggests an elastic flux creep. While with increasing magnetic field, the flux creep becomes faster and system enters into the plastic creep regime. This crossover may correspond to the second peak field, or slightly higher than that field. Finally, the motion of vortices
changes into vortex liquid state above the irreversibility field or temperature. However, a quite large area between the $H_{irr}-T$ and $H_{SP}-T$ curves in the
 intermediate temperature region suggests that the vortex dissipation is still through a plastic motion in this region, via for example the proliferation of the vortex kinky loops\cite{BlatterRMP}. At very high temperatures and magnetic fields, we see the vortex liquid behavior with the very fast vortex motion. We fit the $H_{c2}(T)$ and $H_{irr}(T)$ curves and find that they are well fitted by the expressions
$H_{c2}(T)=H_{c2}(0)(1-T/T_c)^{1.42}$ and $H_{irr}(T)=H_{irr}(0)(1-T/T_c)^{1.51}$. The two lines are very close to each other, which
suggests that the flux liquid region in this material is rather narrow. This is because the material itself is more three dimensional like, not that much two dimensional as in the cuprates.

\section{Concluding remarks}
In conclusion, we have investigated the vortex dynamics through the dynamical and conventional magnetization
relaxation methods on optimally doped Ba$_{0.59}$K$_{0.47}$BiO$_{3+\delta}$ single crystal with $T_c$ = 31.3 K. A second
peak effect is obvious on the magnetization hysteresis loops over a very wide temperature range from 2 to 24 K.
Based on the Bean critical state model and Maley's method, we calculate the transient critical current density and obtained the characteristic pinning energy. We also get the glassy exponent of $\mu\approx$ 1.2 to 1.7 for the magnetic field of 1 T and 3 T in low temperature regime. It indicates that the vortex collective pinning model can describe the vortex dynamics in this material very well.
The second peak position become very close to the irreversibility field at the temperatures above $0.9T_c$, which may
suggest that the vortex phase transition changes from the second order to the first order. Finally, the vortex phase diagram is obtained by combining the resistivity and magnetization measurements. Our results show that the optimally doped BKBO is a model platform to investigate the vortex collective pinning and related physics.

\section*{Acknowledgments}
The authors acknowledge the help of Jian Tao for sharing the experience in growing the samples. This work was supported by the National Key Research and Development Program of China (2016YFA0300401,2016YFA0401700), and the National Natural Science Foundation of China (NSFC) with the projects: A0402/11534005, A0402/11374144.


\begin{thebibliography}{00}
\bibitem{BlatterRMP} G. Blatter, M. V. Feigelman, V. B. Geshkenbein, A. I. Larkin, and V. M. Vinokur, Rev. Mod. Phys. \textbf{66}, 1125 (1994).
\bibitem{WenPhysicaC} H. H. Wen, H. G. Schnack, R. Griessen, B. Dam, J. Rector, Physica C \textbf{241}, 353(1995).
\bibitem{Krusin} L. Krusin-Elbaum, L. Civale, V. M. Vinokur, and F. Holtzberg, Phys. Rev. Lett. \textbf{69}, 2280 (1992).
\bibitem{Klein} L. Klein, E. R. Yacoby, Y. Yeshrun, A. Erb, G. Miiller-Vogt, V. Breit and H. Wiihl, Phys. Rev. B \textbf{49}, 4403 (1994).
\bibitem{PRB72} D. V. Shantsev, A. V. Bobyl, Y. M. Galperin, T. H. Johansen, and S. I. Lee ,Phys. Rev. B \textbf{72}, 024541 (2005).
\bibitem{PRB73} D. V. Denisov, A. L. Rakhmanov, D. V. Shantsev, Y. M. Galperin, and T. H. Johansen Phys. Rev. B \textbf{73}, 014512 (2006).
\bibitem{PRB70} J. P. Rodriguez. Phys. Rev. B \textbf{70}, 224507 (2004).
\bibitem{PRBThompson} J. R. Thompson, Y. R. Sun, D. K. Christen, L. Civale, A. D. Marwick, and F. Holtzberg, Phys. Rev. B \textbf{49}, 13287 (1994).
\bibitem{PRB62} L. Miu, T. Noji, Y. Koike, E. Cimpoiasu, T. Stein, and C. C. Almasan, Phys. Rev. B \textbf{62}, 15172 (2000).
\bibitem{YangHPRB} H. Yang, C. Ren, L. Shan, H. H. Wen, Phys. Rev. B \textbf{78}, 092504 (2008).
\bibitem{NPRB84} N. Haberkorn, M. Miura, B. Maiorov, G. F. Chen, W. Yu, and L. Civale, Phys. Rev. B \textbf{84}, 094522 (2011).
\bibitem{TPRB86} T. Taen, Y. Nakajima, T. Tamegai, and H. Kitamura, Phys. Rev. B \textbf{86}, 094527 (2012).
\bibitem{Bi2212}H. Beidenkopf, N. Avraham, Y. Myasoedov, H. Shtrikman, E. Zeldov, B. Rosenstein, E. H. Brandt, and T. Tamegai, Phys. Rev. Lett. \textbf{95}, 257004 (2005).
\bibitem{Mattheiss} L. F. Mattheiss, E. M. Gyorgy, and D. W. Johnson, Phys. Rev. B \textbf{37}, 3745 (1988).
\bibitem{Cavanature} R. J. Cava, B. Batlogg, J. J. Krajewski, R. Farrow, L. W. Rupp Jr, A. E. White, K. Short, W. F. Peck,
 T. Kometani, Nature (London) \textbf{332}, 814 (1988).
\bibitem{PRB41} S. Y. Pei, J. D. Jorgensen, B. Dabrowski, D. G. Hinks, D. R. Richards, A. W. Mitchell, J. M. Newsam, S. K. Sinha, D. Vaknin, and A. J. Jacobson, Phys. Rev. B \textbf{41}, 4126 (1990).
\bibitem{PRB49} M. Affronte, J. Marcus, C. Escribe-Filippini, A. Sulpice, H. Rakoto, J. M. Broto, J. C. Ousset, S. Askenazy, and A. G. M. Jansen, Phys. Rev. B \textbf{49}, 3502 (1994)
\bibitem{PRB40} W. K. Kwok, U. Welp, G. Crabtree, K. G. Vandervoort, R. Hulsher, Y. Zheng, B. Dabrowski, and L. G. Hinks, Phys. Rev. B \textbf{40}, 9400 (1989).
\bibitem{PRB58} H. J. Kaufmann, Oleg V. Dolgov, and E. K. H. Salje. Phys. Rev. B \textbf{58}, 9479 (1998).
\bibitem{nature335} D. G. Hinks, D. R. Richards, B. Dabrowski, D. T. Marx, A. W. Mitchell, Nature \textbf{335}, 419 (1988).
\bibitem{prl66} C. K. Loong, D. G. Hinks, P. Vashishta, W. Jin, R. K. Kalia, M. H. Degani, D. L. Price, J. D. Jorgensen, B. Dabrowski,
A. W. Mitchell, D. R. Richards, Y. Zheng, Phys. Rev. Lett. \textbf{66}, 3217 (1991).
\bibitem{Js8} M. Braden, W. Reichardt, W. Schmidbauer, A. S. Ivanov, A. Yu. Rumiantsev, J. Superconductivity. \textbf{8}, 595 (1995).
\bibitem{phyC471} H. J. Kang, Y. S. Lee, J. W. Lynn, S. V. Shiryaev, S. N. Barilo, Physica C \textbf{471}, 303 (2011).
\bibitem{TaojianPRB} J. Tao, Q. Deng, H. Yang, Z. H. Wang, X. Y. Zhu, and H. H. Wen, Phys. Rev. B \textbf{91}, 214516 (2015).
\bibitem{Norton} M. L. Norton, Mat. Res. Bull. \textbf{24}, 1391 (1989).
\bibitem{Nishio357} T. Nishio, H. Minami, H. Uwe, Physica C \textbf{357}, 376 (2001).
\bibitem{Anderson} P. W. Anderson, Phys. Rev. Lett. \textbf{9}, 309 (1962).
\bibitem{PRBFeigelman} M. V. Feigel'man and V. M. Vinokur, Phys. Rev. B \textbf{41}, 8986 (1990).
\bibitem{Malozemoff}A.P. Malozemoff and M.P.A. Fisher, Phys. Rev. B \textbf{42},
6784 (1990).
\bibitem{VinokurPRL} M. V. Feigel'man, V. B. Geshkenbein, A. I. Larkin, V. M. Vinokur, Phys. Rev. Lett. \textbf{63}, 2303 (1989).
\bibitem{WenPRL1997}H. H. Wen, A. F. Th. Hoekstra, R. Griessen, S. L. Yan, L. Fang, and M. S. Si, Phys. Rev. Lett. \textbf{79}, 1559 (1997).
\bibitem{WenPhysicaC1998}Hai-Hu Wen, Paul Ziemann, Henri A. Radovan and Thomas Herzog, Physica C \textbf{305},185 (1998).
\bibitem{KimAnderson} P. W. Anderson, Y. B. Kim, Rev. Mod. Phys. \textbf{36}, 39 (1964).
\bibitem{Bean} C. P. Bean, Rev. Mod. Phys. \textbf{36}, 31 (1964).
\bibitem{Kokkaliaris} S. Kokkaliaris, A. A. Zhukov, P. A. J. de Groot, R. Gagnon, L. Taillefer, and T. Wolf, Phys. Rev. B \textbf{61}, 3655 (2000)
\bibitem{Brandt} E. H. Brandt, Rep. Prog. Phys. \textbf{58}, 1465 (1995).
\bibitem{WenAlloys} H. H. Wen, R. Griessen, D. G. de Groot, B. Dam, J. Rector, J. Alloys Compd. \textbf{195}, 427 (1993).
\bibitem{ShenPRB} B. Shen, P. Cheng, Z. S. Wang, L. Fang, C. Ren, L. Shan, and H. H. Wen, Phys. Rev. B \textbf{81}, 014503 (2010).
\bibitem{RMP68} Y. Yeshurun, A. P. Malozemoff, and A. Shaulov, Rev. Mod. Phys. \textbf{68}, 911 (1996).
\bibitem{Maley} M. P. Maley, and J. O. Willis, Phys. Rev. B \textbf{42}, 2639 (1990).
\bibitem{Kwok2}U. Welp, J. A. Fendrich, W. K. Kwok, G. W. Crabtree, and B. W. Veal, Phys. Rev. Lett. \textbf{76}, 4809 (1996).
\bibitem{Kwok} A. M. Petrean, L. M. Paulius, W. K. Kwok, J. A. Fendrich, and G.W. Crabtree, Phys. Rev. Lett. \textbf{84}, 5852 (2000).
\bibitem{Zeldov}B. Khaykovich, E. Zeldov, D. Majer, T. W. Li, P. H. Kes, and M. Konczykowski, Phys. Rev. Lett. \textbf{76}, 2555 (1996).
\bibitem{Yeshurun} B. Kalisky, Y. Myasoedov, A. Shaulov, T. Tamegai, E. Zeldov, and Y. Yeshurun, Phys. Rev. Lett. \textbf{98}, 107001 (2007).

\end{thebibliography}
\end{document}